\documentclass[a4paper,11pt]{article}

\usepackage{jheppub} 
\usepackage{orcidlink}
\usepackage[T1]{fontenc} 

\usepackage{hyperref}
\hypersetup{
    pdfstartview=FitH,
    bookmarksnumbered=true,
    colorlinks,
    breaklinks,
    urlcolor=RoyalBlue,
    linkcolor=RoyalBlue,
    citecolor=NavyBlue
}
\usepackage{graphicx}
\DeclareGraphicsExtensions{.pdf,.png,.jpg,.mps}
\usepackage{booktabs}
\usepackage[dvipsnames]{xcolor}
\usepackage{tikz}
\usetikzlibrary{calc,matrix,decorations.pathmorphing,decorations.markings,arrows,positioning,intersections,mindmap,backgrounds}
\allowdisplaybreaks[4]

\usepackage{amsmath,amsthm,amsfonts,amssymb,amscd,mathtools
}
\usepackage{pifont}
\usepackage{subcaption}

\makeatletter
\newcommand{\ClearSingles}[1]{
  \foreach \k in {1,...,#1}{
    \global\expandafter\let\csname S@\k\endcsname\relax
  }
}
\newcommand{\SetSingles}[2]{
  \ClearSingles{#1}
  \edef\SinglesTmp{#2}
  \foreach \raw in \SinglesTmp {
    \pgfmathtruncatemacro{\s}{\raw}
    \global\expandafter\def\csname S@\s\endcsname{1}
  }
}
\makeatother

\newcommand{\MakeEdgeMidpoints}{
  \foreach \k in {1,...,\N}{
    \pgfmathtruncatemacro{\kp}{mod(\k,\N)+1}%
    \coordinate (E\k) at ($(V\k)!0.5!(V\kp)$);
  }
}

\newcommand{\EdgeXLabels}{
  \foreach \k in {1,...,\N}{
    \node[font=\small, inner sep=0pt]
      at ($(E\k)!-0.35!(0,0)$) {$x_{\k}$};
  }
}

\makeatletter
\newcommand{\ClearEdgeMarks}[1]{
  \foreach \k in {1,...,#1}{
    \global\expandafter\let\csname M@\k\endcsname\relax
  }
}
\newcommand{\SetEdgeMarks}[2]{
  \ClearEdgeMarks{#1}
  \foreach \k/\lab in {#2}{
    \global\expandafter\edef\csname M@\k\endcsname{\lab}
  }
}
\makeatother

\newcommand{\EdgeCustomLabels}{
  \makeatletter
  \foreach \k in {1,...,\N}{
    \expandafter\ifx\csname M@\k\endcsname\relax
      \node[font=\small, inner sep=0pt]
        at ($(E\k)!-0.35!(0,0)$) {$x_{\k}$};
    \else
      \node[font=\small, inner sep=0pt]
        at ($(E\k)!-0.35!(0,0)$) {\csname M@\k\endcsname};
    \fi
  }
  \makeatother
}

\newcommand{\LoopBase}[2]{
  \pgfmathtruncatemacro{\N}{#1}
  \def\R{1.45}
  \def\L{0.85}
  \def\beta{18}
  \pgfmathsetmacro{\thetaZero}{90}
  \SetSingles{\N}{#2}

  \foreach \k in {1,...,\N}{
    \pgfmathsetmacro{\ang}{\thetaZero - 360*(\k-1)/\N}
    \coordinate (V\k) at (\ang:\R);
  }

  \draw (V1) \foreach \k in {2,...,\N}{ -- (V\k) } -- cycle;

  \MakeEdgeMidpoints
  \EdgeXLabels

  \makeatletter
  \foreach \k in {1,...,\N}{
    \pgfmathsetmacro{\ang}{\thetaZero - 360*(\k-1)/\N}
    \expandafter\ifx\csname S@\k\endcsname\relax
      \draw (V\k) --++ (\ang+\beta:\L);
      \draw (V\k) --++ (\ang-\beta:\L);
    \else
      \draw (V\k) --++ (\ang:\L);
    \fi
  }
  \makeatother
}

\newcommand{\LoopBaseCustom}[3]{
  \pgfmathtruncatemacro{\N}{#1}
  \def\R{1.45}
  \def\L{0.85}
  \def\beta{18}
  \pgfmathsetmacro{\thetaZero}{90 + 180/\N}
  \SetSingles{\N}{#2}
  \SetEdgeMarks{\N}{#3}

  \foreach \k in {1,...,\N}{
    \pgfmathsetmacro{\ang}{\thetaZero - 360*(\k-1)/\N}
    \coordinate (V\k) at (\ang:\R);
  }

  \draw (V1) \foreach \k in {2,...,\N}{ -- (V\k) } -- cycle;

  \MakeEdgeMidpoints
  \EdgeCustomLabels

  \makeatletter
  \foreach \k in {1,...,\N}{
    \pgfmathsetmacro{\ang}{\thetaZero - 360*(\k-1)/\N}
    \expandafter\ifx\csname S@\k\endcsname\relax
      \draw (V\k) --++ (\ang+\beta:\L);
      \draw (V\k) --++ (\ang-\beta:\L);
    \else
      \draw (V\k) --++ (\ang:\L);
    \fi
  }
  \makeatother
}

\newcommand{\Diagabdfp}

\title{\Large{Energy Correlators from Star Integrals via Mellin Space }}

\author[a]{Anastasia Volovich\,\orcidlink{0009-0008-2506-3207},}
\author[b]{Di Wu\,\orcidlink{0009-0007-8766-9362}}
\author[c,d]{and Kai Yan\,\orcidlink{0000-0001-8327-7061}}

\emailAdd{anastasia\_volovich@brown.edu}
\emailAdd{stevediwu@zju.edu.cn}
\emailAdd{yan.kai@sjtu.edu.cn}

\affiliation[a]{\footnotesize Department of Physics,
    Brown University,
    Providence,
    RI 02912,
    USA
}

\affiliation[b]{\footnotesize School of Physics, Zhejiang University, Hangzhou, Zhejiang 310058, China }

\affiliation[c]{School of Physics and Astronomy, Shanghai Jiao Tong University, Shanghai 200240, China}
\affiliation[d]{
Key Laboratory for Particle Astrophysics and Cosmology (MOE), Shanghai 200240, China}

\abstract{We explore the Mellin space representation for the collinear limit of $N$-point energy correlators in 
${\cal N}=4$ super-Yang-Mills theory. 
We show that these correlators can be written as integro-differential operators acting on star integrals:
one-loop $n$-gons in $n$ dimensions.
For the three-point energy correlator, we 
obtain the Mellin representation, use it to relate the correlator to the massive box integral, and show
how to solve this relation to match with the expected result.
For the four-point energy correlator, we obtain the Mellin representation and use it to write the
correlator to a sum of various box and hexagon integrals in special kinematics.
Our results provide a systematic method to relate higher-point energy correlators in the collinear limit to star integrals, which are known exactly.}

\begin{document} 

\maketitle

\section{Introduction}
\label{sec:introduction}

In recent decades, our understanding of scattering amplitudes has undergone a revolution, revealing hidden mathematical structures and enabling powerful new calculational tools. At the same time, cross-section–level observables central to collider physics remain comparatively less explored, with the notable exception of energy correlators—an important class of observables that has attracted increasing attention in recent years; see~\cite{Moult:2025nhu} for a comprehensive review by Moult and Zhu.

Energy correlators are of central importance in both collider phenomenology and in the formal study of quantum field theory. They are defined as energy-weighted cross-sections and measure the energy deposited in detectors as a function of the angular separation between detector pairs.   
The study of 
energy correlators 
has a long history in collider physics, where they were initially used to characterize asymptotic energy flux in QCD
~\cite{PhysRevD.17.2298,LOUISBASHAM1979297}. 
The simplest of all collider observables is the one-point energy correlator which was first introduced by Sterman \cite{Sterman:1975xv}
who pointed out that it is infrared finite.
Two decades ago, energy correlators were revisited by Hofman and Maldacena \cite{Hofman:2008ar},
who highlighted them as interesting observables in generic quantum field theories.

Advances in perturbative techniques and conformal field theory have led to significant progress in higher-loop computations of energy correlators, in QCD and in $\mathcal{N}=4$ super-Yang-Mills theory (SYM)~\cite{Belitsky:2013xxa,Belitsky:2013ofa,Dixon:2018qgp,Henn:2019gkr,Dixon:2019uzg,Korchemsky:2019nzm,Kologlu:2019mfz,Chen:2019bpb,Chicherin:2020azt,Yan:2022cye,Yang:2022tgm,  Chicherin:2024ifn,He:2024hbb,Ma:2025qtx,He:2025zbz,Gong:2025jqi,Dempsey:2025yiv},
as well as in pure gravity and $\mathcal{N}=8$ supergravity~\cite{Herrmann:2024yai,Chicherin:2025keq,Ruan:2026xyd}. In particular, the two-point energy correlator has been computed at 
 N$^3$LO in  SYM \cite{Henn:2019gkr} and 
at N$^2$LO in QCD \cite{Dixon:2018qgp}, and even at finite coupling in SYM \cite{Dempsey:2025yiv}.
The three-point energy correlator
has been computed at LO for generic scattering angle in both SYM \cite{Yan:2022cye} and QCD \cite{Yang:2022tgm}.
Higher-point energy correlators have been investigated in their multi-collinear limit and beyond~\cite{Chicherin:2024ifn,Ma:2025qtx,Gong:2025jqi,He:2025zbz,He:2024hbb}. 
In SYM the integrands for the multi-collinear limit of the $N$-point energy correlator have recently been computed 
up to $N=11$ \cite{He:2024hbb}
while the resulting integrals
have been obtained for $N=3$ \cite{Chen:2019bpb} and $N=4$ \cite{Chicherin:2024ifn}. In QCD, analytic results for $N$-point energy correlators are only available for $N=3$ \cite{Chen:2019bpb,Gong:2025jqi}.
Despite remarkable progress, performing the phase-space integrals for higher-point and higher-loop energy correlators remains highly challenging and lacks a general practical algorithm.

In this paper we suggest a new method for studying the collinear limit of $N$-point energy correlators
in SYM by working with their Mellin
representations, which has proven useful in the past in (among other things) the study of dual conformal
integrals.  In particular, using Mellin space, a large class of higher-loop integrals can be written as
simple integro-differential operators acting on star integrals: the one-loop $n$-gons in $n$
dimensions~\cite{Paulos:2012nu,Nandan:2013ip}.  These equations have been crucial, for example,
in evaluating the full set of integrals relevant for two-loop amplitudes in  SYM \cite{Spiering:2024sea}.
Similarly, we will see that working in Mellin space allows one to connect the collinear limit of $N$-point energy correlators
in SYM to star integrals.
These integrals have a beautiful mathematical structure,
being related to volumes of hyperbolic simplices,
and can be computed exactly in terms of
alternating polylogarithms
\cite{Bourjaily:2019exo, Ren:2023tuj}.
This representation
opens the door for using properties and techniques that are well-known in the context of ordinary Feynman
integrals to the more complicated integrals that appear in energy correlators.

This paper is organized as follows. In Section~\ref{sec:two} we review basic facts about the Mellin representation of star integrals.  In Section~\ref{sec:three} we describe how to derive the Mellin representation for a general
multi-point energy correlator in SYM.  In Section~\ref{sec:four} we derive the Mellin
representation of the three-point correlator, write the integro-differential operator that relates it to the
massive box integral, and show that this equation can be readily solved to obtain a formula that matches
previous calculations of this quantity~\cite{Chen:2019bpb}.  In Section~\ref{sec:five} we derive the Mellin representation
of the four-point correlator as a sum of massive box and four hexagon integrals in special kinematics.  In Appendix~\ref{appendixA} we present Mellin representations for hexagon integrals in special kinematics.

\section{Star Integrals in Mellin Space}
\label{sec:two}

In this section, we review the key formulas of \cite{Paulos:2012nu} that will be needed in
the next sections. The one-loop scalar $n$-gon integral in $n$ dimensions, which we will call the \emph{star} integral, referring to its dual graph, is the function of $n$ momenta $P_i$ given by
\begin{equation}
\label{eq:star}
I_n=\pi^{-\frac{n}{2}}\int d^{n}Q\,
\prod_{i=1}^{n}\frac{\Gamma(\Delta_i)}{\bigl(-P_i\!\cdot Q\bigr)^{\Delta_i}},
\end{equation}
where $Q$ is the integration variable and $\Delta_i$ are the propagator powers. It was shown in \cite{Paulos:2012nu} that (\ref{eq:star}) can be expressed in Mellin space as
\begin{equation}
I_n=\oint d\delta_{ij}\;\prod_{i<j}\Gamma(\delta_{ij})\,P_{ij}^{-\delta_{ij}},
\label{eq:ngon-mellin}
\end{equation}
where $P_{ij}\equiv -2\,P_i\cdot P_j$ and the Mellin variables $\delta_{ij}$ are symmetric ($\delta_{ij} = \delta_{ji}$) and constrained to satisfy
\begin{equation}
\label{eq:constraint}
\delta_{ii}=-\Delta_i,\qquad \sum_{j=1}^n\delta_{ij}=0.
\end{equation}
Overall there are $n(n-3)/2$ independent $\delta_{ij}$, which is the same as the number of cross-ratios that one can form with the $P_{ij}$.

For the fully massive (this means all $P_{ij} \ne 0$) box $n=4$, the constraints (\ref{eq:constraint}) can be solved in terms of two variables $\delta_u$, $\delta_v$, and the integral can be expressed explicitly as
\begin{equation}
\label{eq:box}
\hat I_4 \equiv
(P_{13} P_{24}) I_4
=\oint d\delta_u\,d\delta_v\,
\Gamma^2(\delta_u)\Gamma^2(\delta_v)\Gamma^2(1-\delta_u-\delta_v)\,u^{-\delta_u}v^{-\delta_v},
\end{equation}
where 
\begin{equation}
u \equiv \frac{P_{12}P_{34}}{P_{13}P_{24}}\,,\qquad
v \equiv \frac{P_{14}P_{23}}{P_{13}P_{24}}\,.
\end{equation}

For the fully massive hexagon $n=6$, the constraints can be solved in terms of nine independent Mellin variables $\delta_i$ and the integral takes the form
\begin{equation}
\!\!\!\hat I_6\! \equiv\!
(P_{14} P_{25} P_{36}) I_6
\!=\!
\oint \!\prod_{i=1}^{9}d\delta_iu_i^{-\delta_i}\!
\prod_{i=4}^{9}\!\Gamma(\delta_i)\!
\prod_{i=1}^{6}\!\Gamma(\delta_{(i)_3}\!\!-\delta_{(i+1)_6}\!\!-\delta_{(i+2)_6})\!
\prod_{i=1}^{3}\Gamma(1\!-\delta_i-\delta_{(i+1)_3}\!+\delta_{(i+2)_6}\!+\delta_{(i+5)_6}),
\label{hexagon}
\end{equation}
where
\begin{align}
u_i &= u_{i,i+3},~~~\qquad ~~~~~~~~~~~i=1,2,3 \\
u_{i+3} &= u_{i+1,i+5}, ~~~\qquad ~~~~~~~i=1,\ldots,6
\end{align}
in terms of
\begin{equation}
u_{i,j} \equiv \frac{P_{i,j+1}\,P_{i+1,j}}{P_{i,j}\,P_{i+1,j+1}}
\end{equation}
where $(i)_{(n)}$ means $i$ is understood modulo $n$. 

In this paper we will typically encounter integrals in special kinematics, which means that the $P_i$ are not generic but will be subject to various constraints. The key fact we would like to emphasize is that restricting to special kinematics results in an integral with a simpler structure of $\Gamma$-function factors than the original integral. This can happen in two ways. If a kinematic limit involves taking some cross-ratio $u_i \to 0$, this can be implemented at the level of the Mellin representation by computing a residue of the integrand where the corresponding $\delta_i = 0$. On the other hand, if a kinematic limit involves taking some cross ratio $u_i \to 1$, then the corresponding factor $u_i^{-\delta_i}$ in the integrand becomes trivial and typically some of the integrals over the Mellin parameters can be performed analytically using Barnes lemmas or other Mellin integral identities.  In either case, $u_i \to 0$ or $u_i \to 1$, the end result is a Mellin representation with a simpler set of $\Gamma$ factors than the starting point.

\section{Energy Correlators in Mellin Space}
\label{sec:three}

In this section we explain how to write Mellin representations for energy correlators. At leading order, the collinear limit of the $N$-point energy correlator in SYM can be written as a manifestly finite  $(N\!-\!1)$-fold integral over energy fractions $x_1,\ldots,x_N$ of the $1 \to N$ splitting function $\mathcal{G}_N$ as
\cite{Chen:2019bpb,Chicherin:2024ifn,He:2024hbb}
\begin{equation}
\text{E}^N\text{C}=
\frac{1}{z_{12}\cdots z_{N-1N}}
\int \frac{d^{N}x}{\text{GL}(1)}\;
(x_1+ \cdots+x_N)^{-N}\,\mathcal{G}_N
+\text{perm}(1,\ldots,N).
\label{eq:eNc_master}
\end{equation}
The energy correlator $\text{E}^N\text{C}$ is a function of the positions of the $N$ detectors, given by complex numbers $z_i$ representing $N$ points on the sphere. We use 
\begin{equation}
z_{ij}=|z_i-z_j|^2 
\end{equation}
and
the specific and convenient gauge fixing for the
${\text{GL}(1)}$ 
\begin{equation}
\label{eq:measure}
\int \frac{d^{N}x}{\text{GL}(1)}=
\int dx_1 \cdots dx_N \delta (1-\sum_{i=1}^N x_i).
\end{equation}

The explicit expressions for the splitting function $\mathcal{G}_N$ have been given in \cite{Chicherin:2024ifn} for $N=3$ and  $N=4$, and later computed up to $N=11$
in \cite{He:2024hbb}. We will display them explicitly for $N=3$ and $N=4$ in sections 4 and 5, but for now, it is sufficient to discuss their general structure. This will allow us to give a general recipe for constructing a Mellin space representation of $\text{E}^N\text{C}$.

In general $\mathcal{G}_N$ is a linear combination of terms that are monomials in quantities $s_{a\ldots b}$ and $x_{i\ldots j}$ defined by
\begin{equation}
s_{a \ldots b}=\sum_{a \leq i<j\leq b} x_i x_j z_{ij}~~~~~~~~~~~
x_{i \ldots j}=x_i+x_{i+1}+...+x_j.
\end{equation}
We now describe how to write a Mellin representation for a generic term of this type.

{\bf Step 1. } Use the universal factor $X^{-N}$, with $X:=x_1+\cdots+x_N$, to convert the projective measure (\ref{eq:measure}) into a flat measure. This is done by introducing a Schwinger parameter to write
\begin{equation}
\frac{1}{X^N}
=\frac{1}{\Gamma(N)}\int_{0}^{\infty} dt\; t^{N-1}\,e^{-tX}\,.
\label{schw}
\end{equation}
After inserting this representation, one performs the change of variables $x_i'=t x_i$. The homogeneity of the integrand ensures that the Jacobian is compensated by the power of $t$, making the delta function trivial. This leads to
\begin{equation}
    \text{E}^N\text{C} = \frac{1}{z_{12}\cdots z_{N-1N}} 
    \frac{1}{\Gamma(N)}
    \int
    d^N x\, e^{-X}\,\mathcal{G}_N + \text{perm}(1,\ldots,N),
\end{equation}
with no more ${\text{GL}(1)}$ invariance.

{\bf Step 2.} If the denominator has more than one $s_{a..b}$ factor, use a Feynman parameterization to combine them, for example
\begin{equation}
\frac{1}{s_{123}\,s_{234}}=\int_0^\infty d\lambda\;\frac{1}{\bigl(s_{123}+\lambda\,s_{234}\bigr)^2}\,,
\label{feyn}
\end{equation}
and then exponentiate the denominator using \eqref{schw}. 
If the denominator has more than one $x_{i...j}$ factor and they overlap, promote one of them into a quadratic factor using \eqref{feyn} again.

{\bf Step 3.} Mellin transform each exponential factor $x_ix_j z_{ij}$ using the standard formula
\begin{align}
e^{-x_ix_jz_{ij}} = \frac{1}{2 \pi i} \int d\gamma_{ij}\, \Gamma(\gamma_{ij})\, (x_ix_jz_{ij})^{-\gamma_{ij}},
\label{Barnes}
\end{align}
where it is implied that the contour for $\gamma_{ij}$ is parallel to the imaginary axis with ${\rm Re}(\gamma_{ij})>0$.

{\bf Step 4.} At this stage, it is easy to perform all integrals of the $x$'s and the Schwinger and Feynman parameters,
leaving only Mellin integrals, leading to a formula that has the general structure
\begin{align}
    \text{E}^N\text{C} = \sum_a \int [d\gamma_{ij}]_a \, \mathcal{M}_a\, \mathcal{H}_a \prod_{i<j} z_{ij}^{-\gamma_{ij}}
\end{align}
where the $\mathcal{H}_a$ are products of $\Gamma$ functions with arguments that are linear combinations of the $\gamma_{ij}$ and the $\mathcal{M}_a$ are rational functions of the $\gamma_{ij}$ and the
$z_{ij}$.

\section{Three-point Energy Correlator}
\label{sec:four}

In this section we apply the algorithm described in the previous section to derive a Mellin representation for the three-point energy correlator. Then, we show that this representation can be used to derive a simple integro-differential equation between the energy correlator and the box integral (\ref{eq:box}). Finally, we explain how to solve this equation to derive the known expression for the three-point correlator \cite{Chen:2019bpb} starting from the symbol of the box function.

\subsection{Mellin representation}

The three-point splitting function that appears in the starting point (\ref{eq:eNc_master}) is given by 
\cite{Chen:2019bpb}
\begin{align}
{\mathcal{G}}_3=1 + \frac{x_1 x_3}{x_{12} x_{23}} + \frac{s_{12} x_{123}}{s_{123}x_{12}} + \frac{s_{23} x_{123}}{s_{123} x_{23}}.
\label{G3}
\end{align}
Let us start by focusing on the third (or, the first non-trivial) term.  Applying the first two steps to this term gives
\begin{align}
{\rm E^3C}|_{\rm one~term}
& \equiv
\int_{0}^{\infty}\frac{d^3x}{{\rm GL}(1)}\;
\frac{s_{12}}{x_{123}^{2}\,s_{123}\,x_{12}}\\
&=
\int_{0}^{\infty} d^3x\; e^{-x_{123}}\;
\frac{s_{12}}{s_{123}\,x_{12}}\\
&=
\int_{0}^{\infty} d^3x \int_{0}^{\infty} dc_0\,dc_1\;
e^{-x_{123}-c_0 s_{123}-c_1 x_{12}}\; s_{12}\,.
\end{align}
Then we implement step three with
\begin{equation}
e^{-c_0 s_{123}}
=
\oint d\gamma_{12}\,d\gamma_{13}\,d\gamma_{23}\;
\Gamma(\gamma_{12})\Gamma(\gamma_{13})\Gamma(\gamma_{23})\;
c_0^{-(\gamma_{12}+\gamma_{13}+\gamma_{23})}\;
\prod_{i<j}^3(x_i x_j z_{ij})^{-\gamma_{ij}}\,.
\end{equation}
At this stage the $x_i$ integrals factorize and can be done using
\begin{align}
\int_0^\infty dx\, x^{-B} e^{-A x} = A^{B-1} \Gamma(1-B) \qquad \text{for } {\rm Re}(A) > 0 \text{ and } {\rm Re}(B) < 1\,.
\end{align}
The $c_0$ and $c_1$ integrals can be done using
\begin{align}
\int_0^\infty dc_0\, c_0^{a} &= 2 \pi \delta(1+a), \\
\int_0^\infty dc_1\, (1 + c_1)^{-a} &= \frac{1}{a-1}
\qquad \text{for } {\rm Re}(a) > 1\,.
\end{align}
The former is divergent in the common sense, but in the Mellin representation it effectively takes the distributional form shown. Finally, step four gives
\begin{align}
\label{eq:three}
{\rm E^3C}|_{\rm one~term}
=\oint d\gamma_{12}\, d\gamma_{23}
\frac{\gamma_{23}\, (1 {-} \gamma_{12} {-} \gamma_{23})}{2 - \gamma_{12}}
\Gamma(\gamma_{12})^2 \Gamma(\gamma_{23})^2 \Gamma(1{-}\gamma_{12}{-}\gamma_{23})^2
\left(\frac{z_{12}}{z_{13}}\right)^{\!\!-\gamma_{12}+1} \!\left(\frac{z_{23}}{z_{13}}\right)^{-\gamma_{23}}.
\end{align}
Each Mellin contour runs parallel to the imaginary axis, with real parts chosen such that the arguments of all $\Gamma$ functions have positive real values: $\rm{Re}(\gamma_{12})> 0$, ${\rm Re}(\gamma_{23})>0$ and ${\rm Re}(1-\gamma_{12}-\gamma_{23})>0$.

\subsection{Relation to box}

Comparing the three-point energy correlator expression (\ref{eq:three}) to  the box integral (\ref{eq:box}), we identify
\begin{align}
\label{eq:ident}
\delta_u = \gamma_{12},\qquad
\delta_v = \gamma_{23},\\
u = \frac{z_{12}}{z_{13}}, \qquad
v = \frac{z_{23}}{z_{13}}\,.
\end{align}
The expression is now identical to that of the Mellin representation of the box integral (\ref{eq:box}) except for the integrand
factor
\begin{equation}
    \frac{u\, \delta_v(1-\delta_u-\delta_v)}{2-\delta_u}.
\end{equation}
We will now explain how to relate the three-point energy correlator (\ref{eq:three}) to the integro-differential equation of the box integral (\ref{eq:box}).
First, let's shift $\delta_u \rightarrow\delta_u+1$. After factoring out the $\Gamma$-function and power structure, this factor becomes
\begin{equation}
-\frac{\delta_u^2\delta_v}{(1-\delta_u)(\delta_u+\delta_v)}.
\end{equation}
Second, split it as
\begin{equation}
-\frac{\delta_u\delta_v}{1-\delta_u}
+
\frac{\delta_u\delta_v^2}{(1-\delta_u)(\delta_u+\delta_v)},
\end{equation}
and shift $\delta_v \rightarrow\delta_v-1$ only in the second term. Then, the result becomes
\begin{equation}
-\frac{\delta_u(\delta_v+v (1-\delta_u-\delta_v))}{1-\delta_u}.
\end{equation}
Third, deform the contours back to the standard ones, and compute the associated residue contributions. Restoring the $\delta_v$ contour produces no contribution since the residue at $\delta_v =1$ is $0$. Restoring the $\delta_u$ contour produces a contribution at $\delta_u=0$, and the residue is
\begin{equation}
-\oint d\delta_v\;
\bigl(v+(1-v)\delta_v\bigr)\Gamma(\delta_v)^2\Gamma(1-\delta_v)^2v^{-\delta_v}=-1.
\end{equation}
Finally, (\ref{eq:three}) becomes 
\begin{align}
\label{eq:four}
{\rm E^3C}|_{\rm one~term}
=-\oint d\delta_u\, d \delta_v 
\left[
\frac{\delta_u(\delta_v+v (1-\delta_u-\delta_v))}{1-\delta_u}\right]
\Gamma(\delta_u)^2 \Gamma(\delta_v)^2
\Gamma(1{-}\delta_u{-}\delta_v)^2
u^{-\delta_u} v^{-\delta_v}+1\,.
\end{align}
 We now can read off an integro-differential relation between the three-point energy correlator and the box integral as explained in \cite{Paulos:2012nu}.  First of all, numerator factors
in a Mellin representation can be obtained by acting with Euler differential operators
\begin{equation}
\delta_u \leftrightarrow \hat{\partial}_u\equiv-u\partial_u,\qquad
\delta_v \leftrightarrow \hat{\partial}_v\equiv-v\partial_v.
\end{equation}
Second, the denominator factors in (\ref{eq:four}) can be accounted for by introducing a one-fold integral of the box 
\cite{Paulos:2012nu, Nandan:2013ip} defined by

\begin{equation}
\tilde{I}_{4}(u,v) \equiv
\int_{0}^{1}\!dt\; \hat I_4(tu,v).
\label{eq:Dbox_def}
\end{equation}
Altogether, it follows from (\ref{eq:four}) that
\begin{equation}
{\rm E^3C}|_{\rm one~term}
=-\hat{\partial}_u\Big[\hat{\partial}_v
+v\,(1-\hat{\partial}_u-\hat{\partial}_v)\Big]\,
\tilde{I}_4+1\,.
\label{eq:F1dinv}
\end{equation}
Finally,
\begin{equation}
{\rm E^3C}
= \frac{1}{z_{12} z_{23}}\left[ \left(\frac{\pi^2}{6}+1\right)-\hat{\partial}_u\Big[\hat{\partial}_v
+v\,(1-\hat{\partial}_u-\hat{\partial}_v)\Big]\,
\tilde{I}_4(u,v)+(u \leftrightarrow v) \right]  + \text{perm(1,2,3).}
\label{eq:F1dinvfull}
\end{equation}
where the constant term comes from integrating the first two terms in (\ref{G3}). Note that the operation $u \leftrightarrow v$ acts only on the differential operator and the argument of $\tilde{I}_4(u,v)$, not on the constant term.

To summarize this subsection: the formula (\ref{eq:F1dinvfull}) expresses the integro-differential relation between the three-point energy correlator and the standard box integral (\ref{eq:box}) that can be read off from their respective Mellin representations using the methods described in \cite{Paulos:2012nu}.

\subsection{Solution}
\label{sec:solDEI4}

In this subsection we explain a method to solve the relation (\ref{eq:F1dinv}). Specifically, we show how to write a symbol-level solution to this relation starting from the symbol of the box function. In the end we verify that we obtain the correct known formula for the three-point energy correlator from this streamlined approach.

The symbol of the box integral $I_4$ is given by \cite{Goncharov:2010jf, Spradlin:2011wp}
\begin{align}
\mathcal{S}[I_4 (u,v)] = \frac{1}{\Delta} \;\left[  u \otimes \frac{1 -u +v - \Delta}{1-u +v + \Delta} + v \otimes  \frac{1 +u -v - \Delta}{1+u -v + \Delta}  \right] 
\end{align} 
where $\Delta (u,v) =  \sqrt{(1+ u-v)^2 - 4 u }$. It follows from the definition \eqref{eq:Dbox_def} that, at symbol level, we have
\begin{align}\label{eq:Dboxpau}
\partial_u [ u \times {\tilde{I}_4}] = \frac{1}{\Delta} \;\left[  u \otimes \frac{1 -u +v - \Delta}{1-u +v + \Delta} 
+ v \otimes  \frac{1 +u -v - \Delta}{1+u -v + \Delta}  \right] 
\end{align} 
and 
\begin{align}\label{eq:Dboxpav}
\partial_v [ u \times \tilde{I}_4] & = \frac{1-u'-v}{2 v\,  \Delta'} \left[  u' \otimes \frac{1 -u' +v - \Delta'}{1-u' +v + \Delta'} 
+ v \otimes  \frac{1 +u' -v - \Delta'}{1+u' -v + \Delta'}  \right]\Bigg|_{u'=0}^{u'=u}  \notag \\
& \quad - \frac{1}{2 v} \int_{0}^u  \frac{d u'}{u'}  \otimes v  \notag \\
& = \frac{1-u-v}{2 v\,  \Delta} \left[  u \otimes \frac{1 -u +v - \Delta}{1-u +v + \Delta} 
+ v \otimes  \frac{1 +u -v - \Delta}{1+u -v + \Delta}  \right] \notag \\
& \quad + \frac{1}{2 v}\, \left[ [2] \,  v \otimes (1-v) - u \otimes v - v \otimes u \right] .
\end{align} 
The solution to the first equation \eqref{eq:Dboxpau} must take the form
\begin{align}
\label{eq:step}
\mathcal{S}[u \times \tilde{I}_4 ]& =\frac12\, \left[  u \otimes \frac{1 -u +v - \Delta}{1-u +v + \Delta} 
+ v \otimes  \frac{1 +u -v - \Delta}{1+u -v + \Delta}  \right] \otimes \frac{1-u +v -\Delta}{1-u +v + \Delta}  + \overline{S}(u,v)
\end{align} 
where $\overline{S}$ is any homogeneous solution $\partial_u \overline{S} = 0$.
Substituting (\ref{eq:step}) into the second equation \eqref{eq:Dboxpav} shows that the $\overline{S}$ must satisfy
\begin{align}
\partial_v \, \overline{S} = \frac{1}{2 v}\, \left[[2]\,  v \otimes (1-v) - u \otimes v - v \otimes u \right] \,. 
\end{align} 
which suggests the solution
\begin{align}
\overline{S} =\frac12 \, \left[[2] \, v \otimes (1-v) - u \otimes v - v \otimes u \right] \otimes v.
\end{align} 
Putting everything together, we obtain the symbol-level solution
\begin{align}\label{eq:SDbox}
\mathcal{S}[ \tilde{I}_4] &  = \frac{1}{2u}  \left[  u \otimes \frac{1 -u +v - \Delta}{1-u +v + \Delta}  \otimes \frac{1-u +v -\Delta}{1-u +v + \Delta} 
+ v \otimes  \frac{1 +u -v - \Delta}{1+u -v + \Delta}   \otimes \frac{1-u +v -\Delta}{1-u +v + \Delta}  \right. \notag \\
& + [2]\, v \otimes (1-v) \otimes v  - u \otimes v \otimes v - v \otimes u  \otimes v \Big].
\end{align}
Plugging \eqref{eq:SDbox} into \eqref{eq:F1dinv}, we can work out its symbol and compare with the result given in \cite{Chen:2019bpb} (or by direct integration).  The answer is expressed in terms of two pure weight-two functions
\begin{align}
D_{-} \equiv 2\, \text{Li}_2 (z)- 2 \, \text{Li}_2 (\bar z ) + \ln |z|^2 \ln \frac{1-z}{1-\bar z} \,, \quad  D_{+} \equiv \text{Li}_2 (1-v) + \frac{1}{2} \ln v \ln u \,.  
\end{align} 
where $z, \bar{z}$ are related to $u, v$ by $u = z \bar{z}$, $v = (1-z)(1-\bar{z})$; their symbols are given by
\begin{align}
\mathcal{S}[ D_{-}] & =     u \otimes \frac{1 -u +v - \Delta}{1-u +v + \Delta} + v \otimes  \frac{1 +u -v - \Delta}{1+u -v + \Delta},   \\
\mathcal{S}[ D_{+}] & =  [-1]\,  v \otimes (1-v) + \big[\frac12 \big] \, v \otimes u + \big[\frac12 \big]\, u \otimes v. 
\end{align} 
Hence, at the symbol level, the following equations hold, which follow from \eqref{eq:Dboxpau} and \eqref{eq:Dboxpav} 
\begin{align}
& (1- \hat{\partial}_u) \, \tilde{I}_4  = \frac{1}{\Delta} \, D_{-}  \\  
& \hat{\partial}_v \, \tilde{I}_4  =  \frac{1-u-v}{2 u\, \Delta}\,  D_{-} - \frac{1}{u}\, D_{+}    
\end{align} 
Finally, recalling \eqref{eq:F1dinv},  
${\rm E^3C}|_{\rm one~term}$
can be expressed in terms of $D_{+}, D_{-}$ and their first derivatives  as 
\begin{align}
 {\rm E^3C}|_{\rm one~term}&= u\, \partial_u \left[  \frac{-1 + u + 2 v + u v - v^2}{2 u\, \Delta} D_{-}   + \frac{1-v}{u} D_{+} \right]+1.
\end{align}
Putting everything together, we arrive at the final answer
{\small{
\begin{align}
    {\rm E^3C} = \frac{1}{z_{12} z_{23}}\left[ \left(\frac{\pi^2}{6}+1\right) + u \partial_u \left[  \frac{-1 + u + 2 v + u v - v^2}{2 u\, \Delta} D_{-}   + \frac{1-v}{u} D_{+} \right] + (u \leftrightarrow v)\right]+{\rm perm}(1,2,3)
\end{align}
}}
which agrees with \cite{Chen:2019bpb}.

\section{Four-point Energy Correlator}
\label{sec:five}

In this section we apply the algorithm described Section~\ref{sec:three} to derive a Mellin representation for one particular term in the four-point energy correlator, as an illustrative example of the general method. We explicitly write the integro-differential relation between this term and the hexagon integral (\ref{hexagon}).

\subsection{Mellin representation}

The four-point splitting function is given by \cite{Chicherin:2024ifn}
\begin{equation}
\resizebox{\textwidth}{!}{$
\begin{aligned}
\mathcal{G}_4 =\;&
1+\frac{x_1 x_{34}}{x_{12} x_{234}}
+ \frac{x_4 x_{12}}{x_{34} x_{123}}
+ \frac{s_{12} x_{123}}{s_{123} x_{12}}
+ \frac{s_{34} x_{234}}{s_{234} x_{34}}
+ \frac{s_{123} x_{1234}}{s_{1234} x_{123}}
+ \frac{s_{234} x_{1234}}{s_{1234} x_{234}}
+ \frac{s_{23} s_{1234}}{s_{123} s_{234}}
+ \frac{s_{12} s_{23} x_{1234}^{2}}{s_{234} s_{1234} x_{12} x_{123}}
\\
&+
\frac{s_{23} s_{34} x_{1234}^{2}}{s_{123} s_{1234} x_{34} x_{234}}
+ \frac{x_1 s_{34}^{2} x_{1234}}{s_{234} s_{1234} x_{12} x_{34}}
+ \frac{x_4 x_1^{2} s_{34}}{s_{234} x_{12} x_{123} x_{234}}
+ \frac{x_4^{2} x_1 s_{12}}{s_{123} x_{34} x_{123} x_{234}}
+ \frac{x_4 s_{12}^{2} x_{1234}}{s_{123} s_{1234} x_{12} x_{34}}
\\
&+
\frac{s_{23}^{2} x_{1234}^{2}}{s_{123} s_{234} x_{123} x_{234}}
+ \frac{x_4 x_1 s_{23} s_{1234}}{s_{123} s_{234} x_{123} x_{234}}
+ \frac{x_1 s_{23} x_{34} x_{1234}}{s_{234} x_{12} x_{123} x_{234}}
+ \frac{x_4 s_{23} x_{12} x_{1234}}{s_{123} x_{34} x_{123} x_{234}}
+ \frac{x_4 s_{12} s_{234} x_{1234}}{s_{123} s_{1234} x_{34} x_{234}}
\\[6pt]
&+
\frac{x_1 s_{34} s_{123} x_{1234}}{s_{234} s_{1234} x_{12} x_{123}}
+ \frac{s_{12} s_{34} x_{123} x_{1234}}{s_{123} s_{1234} x_{12} x_{34}}
+ \frac{s_{12} s_{34} x_{234} x_{1234}}{s_{234} s_{1234} x_{12} x_{34}} .
\end{aligned}
$}
\label{Neq4int}
\end{equation}

\begin{figure}
\centering
\begin{subfigure}{0.3\textwidth}
  \centering
  \vspace*{-10em}
\includegraphics[width=0.65\linewidth]{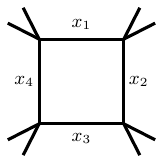}
   \caption{}
\end{subfigure}\hfill
\begin{subfigure}{0.3\textwidth}
  \centering
  \includegraphics[width=0.7\linewidth]{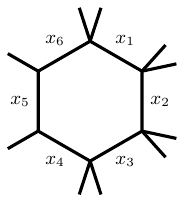}
 \caption{}
\end{subfigure}\hfill
\begin{subfigure}{0.3\textwidth}
  \centering
  \includegraphics[width=0.7\linewidth]{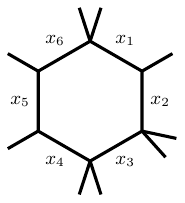}
 \caption{}
\end{subfigure}

\begin{subfigure}{0.45\textwidth}
  \centering
  \includegraphics[width=0.5\linewidth]{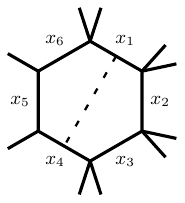}
 \caption{}
\end{subfigure}\hfill
\begin{subfigure}{0.45\textwidth}
  \centering
  \includegraphics[width=0.5\linewidth]{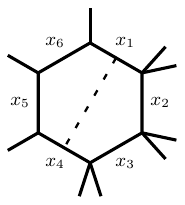}
  \caption{}
\end{subfigure}

\caption{Star integrals corresponding to the five types of $\Gamma$ function structures in $\text{E}^\text{4}\text{C}$. 
(1) 4-mass box.
(2) 4-mass hexagon: $u_2=u_9=1$ and $P_{45}=P_{56}=0$. 
(3) 3-mass hexagon: same as (2), additionally with $P_{12}=0$. 
(4) 4-mass hexagon: given by the integral shown in (\ref{eq:deformedhexagon}), with kinematics
satisfying (\ref{eq:type4constraints}). Points connected with dashed lines are null-separated: $P_{14} = 0$.
(5) 3-mass hexagon: the same integral (\ref{eq:deformedhexagon}), additionally with $P_{16}=0$. } 
\label{fig:hextype}
\end{figure}

Applying the steps described in Section~\ref{sec:three} for integral (\ref{Neq4int}) we obtain a Mellin representation 
\begin{equation}
{\rm E^4C=}
\sum_{a=1}^{5} \int\left[d\gamma_{ij}\right]_a\,\mathcal{M}_a\,\mathcal{H}_a\, \prod_{i<j}  z_{ij}^{-\gamma_{ij}}
\label{eq:G4main}
\end{equation}
with five different types of structures, given explicitly by the following expressions:
{\scriptsize
\begin{align}
\mathcal{H}_1
&=
\Gamma(\delta_u)^2\Gamma(\delta_v)^2\Gamma(1-\delta_u-\delta_v)^2
\\[4pt]
\label{eq:H2def}
\mathcal{H}_2 
&=
-\Gamma(\delta_1)\Gamma(\delta_3-1)\Gamma(\delta_4)\Gamma(\delta_5)\Gamma(\delta_6)\,
\Gamma(\delta_3-\delta_4-\delta_5)\,
\Gamma(1-\delta_1-\delta_3+\delta_5)
\Gamma(1-\delta_3+\delta_4-\delta_6)\,
\Gamma(\delta_1-\delta_5-\delta_6)\,
\Gamma(1-\delta_1-\delta_4+\delta_6) .
\\[4pt]
\mathcal{H}_3
&=
-\Gamma(\delta_1)\,\Gamma(\delta_3-1)\,
\Gamma(\delta_5)\,\Gamma(\delta_6)\,
\Gamma(\delta_3-\delta_5)
\Gamma(1-\delta_1-\delta_3+\delta_5)\,
\Gamma(1-\delta_3-\delta_6)\,
\Gamma(\delta_1-\delta_5-\delta_6)\,
\Gamma(1-\delta_1+\delta_6).
\\[6pt]
\mathcal{H}_4
\label{eq:H4}
&=
-\Gamma(\delta_1)\Gamma(1+\delta_1)\,
\Gamma(\delta_3-1)^2\,
\Gamma(-\delta_1-\delta_4)\Gamma(\delta_4)\,
\Gamma(-\delta_3-\delta_6)\,
\Gamma(1-\delta_3+\delta_4-\delta_6)\,
\Gamma(\delta_6)\,
\Gamma(1-\delta_1-\delta_4+\delta_6).
\\[8pt]
\mathcal{H}_5
&=
\frac{
\Gamma(\delta_{1})\,\Gamma(1-\delta_{1})\,
\Gamma(\delta_{3})\,\Gamma(1-\delta_{3})\,
\Gamma(1+\delta_{1})\,\Gamma(\delta_{3}-1)
}{
\Gamma(2-\delta_{1}-\delta_{3})
}
\Gamma(-\delta_{1}-\delta_{4})\,\Gamma(\delta_{4})\,
\Gamma(-\delta_{3}-\delta_{6})\,\Gamma(1-\delta_{3}+\delta_{4}-\delta_{6})\,
\Gamma(\delta_{6})\,\Gamma(1-\delta_{1}-\delta_{4}+\delta_{6}) .
\label{ExprH}
\end{align}
}
These structures of $\Gamma$-functions indicate that these five contributions correspond, respectively, to star integrals of the types shown in Figure~\ref{fig:hextype}, sometimes in special kinematics and sometimes with an additional one-fold integral, as indicated in the figure.

We provide the full expressions for the five  Mellin kernels $\mathcal{M}_a$ and integration measures $[d \gamma_{ij}]_a$ in an ancillary file.
Here let us proceed by analyzing just a single term as an example.
Consider the term

\begin{align}
{\rm E^4C}|_{\rm one ~term}=
\int_0^\infty \frac{d^4 x}{\mathrm{GL}(1)}\frac{s_{23}^2}{s_{123}\,s_{234}\,x_{1234}^{4}}\,.
\label{EC4oneterm}
\end{align}
Using the steps described in Section~\ref{sec:three} we can obtain its Mellin representation 
\begin{align}
\label{eq:four-point-term}
{\rm E^4C}|_{\rm one ~term}=\frac{1}{6}
\oint 
d\gamma_{12}\,d\gamma_{13}\,d\gamma_{24}\,d\gamma_{34}\;
 \,&(\gamma_{12}+\gamma_{13})\,(\gamma_{24}+\gamma_{34})^{2}\,(1+\gamma_{24}+\gamma_{34})^{2}\,
 \mathcal{H}_4(\gamma)\,\\
&\times z_{12}^{-\gamma_{12}}\,z_{13}^{-\gamma_{13}}\,
z_{23}^{\,\gamma_{12}+\gamma_{13}+\gamma_{24}+\gamma_{34}}\,
z_{24}^{-\gamma_{24}}\,z_{34}^{-\gamma_{34}} \, ,
\end{align}
where $\mathcal{H}_4(\gamma)$
is given in (\ref{eq:H4}) with the following substitution
\begin{equation}
\label{delta}
\delta_1=-\gamma_{12}-\gamma_{13},
\qquad
\delta_3=-\gamma_{24}-\gamma_{34},
\qquad
\delta_4=\gamma_{13},
\qquad
\delta_6=\gamma_{24}.
\end{equation}

\subsection{Relation to hexagon}

In this subsection we will show that the term we have focused on in (\ref{EC4oneterm}) can be related to a hexagon integral in special kinematics.
To that end, let us start by defining a one-fold integral of the hexagon 

\begin{equation}
\label{eq:deformedhexagon}
\tilde{I}_6(X_1,X_2,X_3,X_4)\equiv\int_0^\infty 
dt\,
\hat I_6'((1+t){t}^{-1} \,X_1,(1+t)\,X_2,\,X_3,X_4),
\end{equation}
where ${\hat I'_6}$ is the hexagon integral (\ref{hexagon}) evaluated in special kinematics given by imposing the constraints
\begin{equation}
\label{eq:type4constraints}
P_{14}=P_{45}=P_{56}=0,\qquad
P_{26}P_{35}=P_{25}P_{36},\qquad
P_{16}P_{25}=P_{15}P_{26}.
\end{equation}
The last two of these correspond to $u_2 = u_9 = 1$.
After imposing these relations, only four independent cross-ratios remain of the original nine, which we reorganize as
\begin{equation}
X_1 \equiv u_1u_5=\frac{P_{15}P_{23}}{P_{13}P_{25}}, ~
X_2 \equiv u_3u_5=\frac{P_{23}P_{46}}{P_{24}P_{36}}, ~
X_3 \equiv u_4=\frac{P_{12}P_{36}}{P_{26}P_{13}}, ~
X_4 \equiv u_6=\frac{P_{34}P_{25}}{P_{24}P_{35}}.
\end{equation}

The purpose of introducing this specific integral is that one can check that its Mellin representation is 
\begin{align}
\tilde{I}_6 =
\oint d\delta_1\,d\delta_3\,d\delta_4\,d\delta_6\;
X_1^{-\delta_1}\,X_2^{-\delta_3}\,X_3^{-\delta_4}\,X_4^{-\delta_6}\,
\mathcal{H}_4(\delta_1,\delta_3,\delta_4,\delta_6),
\end{align}
which precisely matches that of (\ref{eq:H4}).

The prefactor of \eqref{eq:four-point-term} can be accounted for via a differential operator as described above.  Finally, we conclude that
\begin{equation}
{\rm E^4C}|_{\rm one ~term}
=\frac{1}{6}
\hat{\partial}_1\,\hat{\partial}_2^{\,2}\,\bigl(1-\hat{\partial}_2\bigr)^2\,
\tilde{I}_6
\end{equation}
where $\hat{\partial}_i=-X_i \partial_{X_i}.$

In Appendix A, we will show how the kernels $\mathcal {H} _2$, $\mathcal {H} _3$, and $\mathcal {H} _5$ arise from the fully massive hexagon in special kinematics.

\section{Outlook}

In this paper, we have shown that the collinear limit of $N$-point energy correlators in SYM can be efficiently computed by relating them to simpler, well-understood star integrals via Mellin space.
Specifically, we provide a general computational framework to derive integro-differential relations between $N$-point energy correlators and star integrals. For $N=3$, the energy correlator is related to the box integral, while for $N=4$ it can be written as a sum of boxes and hexagons in special kinematics.
Our method opens the door to higher-point calculations and
to applying amplitude technology, including symbol integration, elliptical integrals and bootstrap methods
to the realm of energy correlators.

We have seen that (just as in the original work~\cite{Paulos:2012nu,Nandan:2013ip}), working in
Mellin space naturally leads
to one-fold (or, in general, higher-fold) integrals of the type encountered in \eqref{eq:Dbox_def}, \eqref{eq:deformedhexagon}. These integrals exhibit distinct features compared with the star integrals themselves. As shown in \cite{Chicherin:2024ifn},  we expect to see 
cubic-root letters in the symbol of the one-fold integral over hexagon defined in \eqref{eq:deformedhexagon}. It would be very interesting to develop a general technology to compute the symbols
of these types of integrals, perhaps along the lines of method used in~\cite{He:2023qld}.  It is also promising to develop an algorithmic approach to construct (canonical) differential equations for energy correlators in Mellin space bypassing integration by parts, as we did for the $N=3$ case in this paper. 

 It would be fascinating to rewrite the 51-term expression found in \cite{Chicherin:2024ifn} 
 as compactly as possible, in the form of integro-differential operators acting on star integrals using the expression we
 obtained in (\ref{eq:G4main}). Star integrals have uniform transcendentality, and a beautifully simple mathematical
 structure (see \cite{Bourjaily:2019exo,Ren:2023tuj} and references therein) -- they are related to volumes of orthoschemes -- so such a representation
 would manifest that all of the 
 non-uniform transcendentality terms
 arise from the operators.

 It is important to better understand the structure of the higher-$N$ correlators, which were computed at the integrand level up to $N=11$ in \cite{He:2024hbb}. Already in the first unknown case $N=5$, the (integrated) energy correlator involves elliptic polylogarithm functions with many distinct elliptic curves, and even more complicated curves and higher-dimensional varieties will appear for $N>5$.
Finally, it will be interesting to see if the methods used in this paper can be helpful for studying energy correlators at higher loop orders and other theories, including QCD and gravity.

\acknowledgments

We would like to thank Jianyu Gong, Ian Moult,
Oliver Schnetz, Marcos Skowronek, Marcus Spradlin, Yichao Tang, Qinglin Yang and Jiarong Zhang for valuable discussions. This work was supported in part by the US Department of Energy under contract DE-SC0010010 Task F and by Simons Investigator Award \#376208 (AV). The work of KY is supported by National Natural Science Foundation of China under Grant No. 12357077.

\appendix

\section{Mellin Representation: Hexagons}
\label{appendixA}

In this appendix we show how the kernels $\mathcal H_2$, $\mathcal H_3$ and $\mathcal H_5$ arise from the fully massive hexagon in special kinematics. The derivation of $\mathcal H_4$ was given in the main text.
\subsection{\texorpdfstring{$\mathcal H_2$}{H2}}
Start from the fully massive hexagon (\ref{hexagon}) and impose
\begin{equation}
P_{45}=P_{56}=0,
\qquad
P_{26}P_{35}=P_{25}P_{36},
\qquad
P_{16}P_{25}=P_{15}P_{26},
\end{equation}
which is equivalent to
\begin{equation}
u_7=u_8=0,
\qquad
u_2=u_9=1.
\end{equation}
In this configuration the remaining independent cross-ratios are $u_1,u_3,u_4,u_5,u_6$. Recall that taking a cross-ratio $u_i \to 0$ in Mellin space can be implemented by evaluating the corresponding residue of the integrand. In practice, this amounts to dropping the associated Mellin integral together with the factor $\Gamma(\delta_i)$, and then setting $\delta_i=0$ in the integrand. The Mellin representation therefore, becomes
\begin{align}
\hat I_6^{(2)}
&=
\oint d\delta_1\,d\delta_2\,d\delta_3\,d\delta_4\,d\delta_5\,d\delta_6\,d\delta_9\;
u_1^{-\delta_1}u_3^{-\delta_3}u_4^{-\delta_4}u_5^{-\delta_5}u_6^{-\delta_6}
\nonumber\\
&\quad\times
\Gamma(\delta_4)\Gamma(\delta_5)\Gamma(\delta_6)\Gamma(\delta_9)\,
\Gamma(\delta_1-\delta_5-\delta_6)\Gamma(\delta_2-\delta_6)\,
\Gamma(1-\delta_1-\delta_3+\delta_5)\Gamma(\delta_3)
\nonumber\\
&\quad\times
\Gamma(\delta_1-\delta_9)\Gamma(1-\delta_1-\delta_2+\delta_6+\delta_9)\,
\Gamma(1-\delta_3-\delta_2+\delta_4)\,
\Gamma(\delta_3-\delta_4-\delta_5)\Gamma(\delta_2-\delta_4-\delta_9).
\end{align}
Note that the variables $\delta_2$ and $\delta_9$ now appear only in a Barnes integral. Integrating over $\delta_2$ and $\delta_9$ using Barnes' first lemma, the $\Gamma$-functions in the denominator cancel, and we obtain
\begin{align}
\hat I_6^{(2)}
=
\oint d\delta_1\,d\delta_3\,d\delta_4\,d\delta_5\,d\delta_6\;
u_1^{-\delta_1}u_3^{-\delta_3}u_4^{-\delta_4}u_5^{-\delta_5}u_6^{-\delta_6}\,
\mathcal H_2(\delta_1,\delta_3,\delta_4,\delta_5,\delta_6),
\end{align}
with $\mathcal H_2$ given in (\ref{eq:H2def}). To identify with the ${\rm E^4C}$ $\Gamma$-function structure, one uses the map
\begin{equation}
\label{MapType2}
\delta_1=\gamma_{34},
\qquad
\delta_3=\gamma_{24},
\qquad
\delta_4=\gamma_{13},
\qquad
\delta_5=\gamma_{34}+\gamma_{24}+\gamma_{23},
\qquad
\delta_6=-\gamma_{24}-\gamma_{23}-\gamma_{12}.
\end{equation}

\subsection{\texorpdfstring{$\mathcal H_3$}{H3}}

For $\mathcal H_3$, together with the constraints for $\mathcal H_2$, we further impose
\begin{equation}
P_{12}=0,
\end{equation}
which sets $u_4=0$. In Mellin space this removes $\Gamma(\delta_4)u_4^{-\delta_4}$ and sets $\delta_4=0$. One obtains
\begin{align}
\hat I_6^{(3)}
=
\oint d\delta_1\,d\delta_3\,d\delta_5\,d\delta_6\;
u_1^{-\delta_1}u_3^{-\delta_3}u_5^{-\delta_5}u_6^{-\delta_6}\,
\mathcal H_3(\delta_1,\delta_3,\delta_5,\delta_6),
\end{align}
and the identification is
\begin{equation}
\label{MapType3}
\delta_1=\gamma_{13}-\gamma_{25},
\qquad
\delta_3=-\gamma_{12}-\gamma_{13},
\qquad
\delta_5=-\gamma_{25}-\gamma_{35},
\qquad
\delta_6=\gamma_{13}.
\end{equation}

\subsection{\texorpdfstring{$\mathcal H_5$}{H5}}

For $\mathcal H_5$, impose
\begin{equation}
P_{45}=P_{56}=P_{16}=P_{14}=0,
\qquad
P_{26}P_{35}=P_{25}P_{36}.
\end{equation}
As in $\mathcal H_4$ case, the original cross-ratios $u_1$, $u_3$ and $u_5$ are then not finite, so we reorganize them as
\begin{equation}
Y_1 \equiv u_1u_5=\frac{P_{15}P_{23}}{P_{13}P_{25}},
\qquad
Y_2 \equiv u_3u_5=\frac{P_{23}P_{46}}{P_{24}P_{36}},
\qquad
Y_3 \equiv u_4=\frac{P_{12}P_{36}}{P_{26}P_{13}},
\qquad
Y_4 \equiv u_6=\frac{P_{34}P_{25}}{P_{24}P_{35}}.
\end{equation}
Removing the three vanishing Mellin factors and performing the Barnes integral over $\delta_2$ gives a four-fold Mellin representation. 
We then define the one-fold integration
\begin{equation}
\tilde I_6^{(5)}(Y_1,Y_2,Y_3,Y_4)\equiv\int_0^\infty 
dt\,
\hat I_6^{(5)\prime} ((1+t){t}^{-1} \,Y_1,(1+t)\,Y_2,\,Y_3,Y_4),
\end{equation}
where $\hat I_6^{(5)\prime}$ denotes the degenerate hexagon. As a result,
\begin{align}
\tilde I_6^{(5)}
=
\oint d\delta_1\,d\delta_3\,d\delta_4\,d\delta_6\;
Y_1^{-\delta_1}Y_2^{-\delta_3}Y_3^{-\delta_4}Y_4^{-\delta_6}\,
\mathcal H_5(\delta_1,\delta_3,\delta_4,\delta_6),
\end{align}
and the identification is
\begin{equation}
\label{MapType5}
\delta_1=-\gamma_{12}-\gamma_{13},
\qquad
\delta_3=-\gamma_{24}-\gamma_{34},
\qquad
\delta_4=\gamma_{13},
\qquad
\delta_6=\gamma_{24}.
\end{equation}

\vspace{6pt}
\vspace{12pt} 
\bibliography{ref}
\bibliographystyle{JHEP}

\end{document}